\documentclass[%
 aip,
 amsmath,amssymb,
reprint,%
]{revtex4-1}

\DeclareRobustCommand{\ez}[1]{ {\begingroup\sethlcolor{BurntOrange}\hl{}\endgroup} }

\usepackage{graphicx}
\usepackage{dcolumn}
\usepackage{bm}
\usepackage{xcolor, soul}
\sethlcolor{green}

\usepackage[utf8]{inputenc}
\usepackage[T1]{fontenc}
\usepackage{mathptmx}
\usepackage{etoolbox}

\makeatletter
\def\@email#1#2{%
 \endgroup
 \patchcmd{\titleblock@produce}
  {\frontmatter@RRAPformat}
  {\frontmatter@RRAPformat{\produce@RRAP{*#1\href{mailto:#2}{#2}}}\frontmatter@RRAPformat}
  {}{}
}%
\makeatother
\begin{document}


\title[]{Atomic step disorder on polycrystalline surfaces leads to spatially inhomogeneous work functions}
\author{Morgann Berg}
\author{Sean W. Smith}
\author{David A. Scrymgeour}
\author{Michael T. Brumbach}
\author{Ping Lu}
\author{Sara M. Dickens}
\author{Joseph R. Michael}
\author{Taisuke Ohta}
\author{Ezra Bussmann$^{*}$}
 \email{ebussma@sandia.gov}
\author{Harold P. Hjalmarson}
\author{Peter A. Schultz}
\author{Paul G. Clem}
\author{Matthew M. Hopkins}
\author{Christopher H. Moore$^{*}$}
 \email{chmoore@sandia.gov}
\affiliation{$^{1}$Sandia National Laboratories, Albuquerque NM, 87185 USA}

\date{\today}

\begin{abstract}

Structural disorder causes materials surface electronic properties, e.g. work function ($\phi$) to vary spatially, yet it is challenging to prove exact causal relationships to underlying ensemble disorder, e.g. roughness or granularity. For polycrystalline Pt, nanoscale resolution photoemission threshold mapping reveals a spatially varying $\phi= 5.70\pm 0.03$~eV over a distribution of (111) textured vicinal grain surfaces prepared by sputter deposition and annealing. With regard to field emission and related phenomena, e.g. vacuum arc initiation, a salient feature of the $\phi$ distribution is that it is skewed with a long tail to values down to 5.4 eV, i.e. far below the mean, which is exponentially impactful to field emission via the Fowler-Nordheim relation. We show that the $\phi$ spatial variation and distribution can be explained by ensemble variations of granular tilts and surface slopes via a Smoluchowski smoothing model wherein local $\phi$ variations result from spatially varying densities of electric dipole moments, intrinsic to atomic steps, that locally modify $\phi$. Atomic step-terrace structure is confirmed with scanning tunneling microscopy (STM) at several locations on our surfaces, and prior works showed STM evidence for atomic step dipoles at various metal surfaces. From our model, we find an atomic step edge dipole $\mu=0.12$ D/edge atom, which is comparable to values reported in studies that utilized other methods and materials. Our results elucidate a connection between macroscopic $\phi$ and nanostructure that may contribute to the spread of reported $\phi$ for Pt and other surfaces, and may be useful toward more complete descriptions of polycrystalline metals in models of field emission and other related vacuum electronics phenomena, e.g. arc initiation.

\end{abstract}

\maketitle

\section{\label{sec:level1}Introduction}
Field emission (FE) of electrons from solid surfaces is an integral mechanism to many electronic technologies, and a fundamental contributor to complex phenomena, e.g. vacuum arcs.~\cite{jen18, box96, tan16}  In cold FE, electron fluxes, $J$, depend on a surface's work function $\phi$, via the Fowler-Nordheim relation $J$~($\phi$, $\beta E$)~$=(A \beta^{2} E^{2} /\phi)$~exp~$(-B \phi^{3/2}/\beta E)$, where $\beta E$ is the local effective electric field, and $A$ and $B$ are constants.~\cite{fow28, for07,jen19}  For realistic surfaces, $\phi$, and consequently $J$, vary spatially owing to near-surface disorder.~\cite{for01,le90,van92,smo41,str73,bes77,haa69,sch21}  Since $J$ depends exponentially on local $\phi$, the consequences of spatially patchy $\phi$ distributions are amplified, and especially impactful for feedback-driven phenomena like vacuum arc initiation.

\begin{figure}
\includegraphics[width=0.4\textwidth]{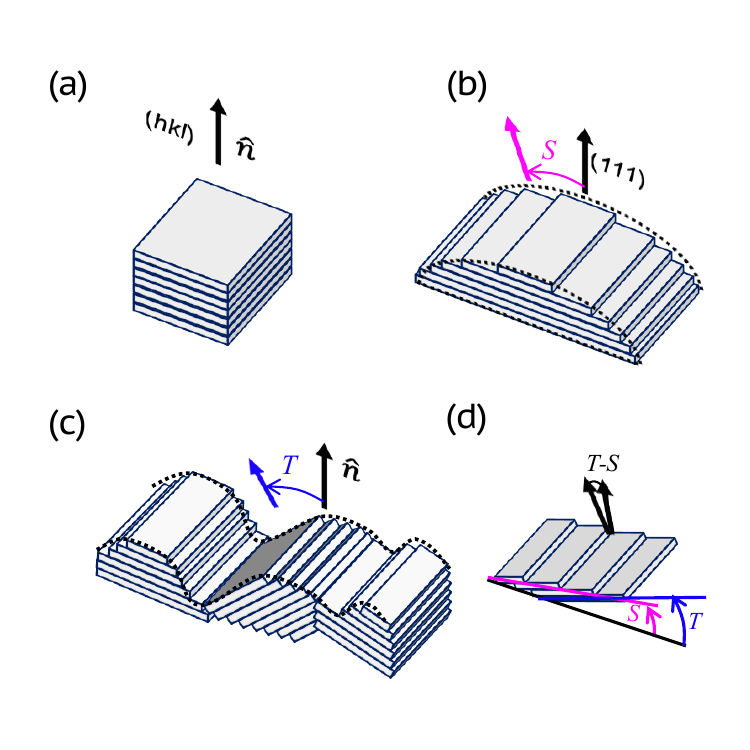}
\caption{\label{fig:fig_1} Schematic relationships between atomic steps and crystal surfaces with increasing complexity. (a) An oriented monocrystalline surface or step-free (hkl) facet. (b) A curved sample with varying slope, $S$, used by Besocke et al. to investigate relationships between atomic step density and work function. (c) The polycrystalline surfaces in our study have locally varying slopes and nonzero varying tilt, $T$, angles with respect to a global (mean) surface normal, $\hat{n}$. (d) In a surface region with slope and tilt, the atomic step density is proportional to the difference between $S$ and $T$.}
\end{figure}

Some $\phi$ spatial variation across crystalline, and polycrystalline, metal surfaces arises from inherent crystal anisotropy effects combined with roughness likely to be present in any realistic scenario.  A crystal's work function is facet (hkl)-dependent, i.e. $\phi = \phi$~(hkl).~\cite{smo41} For some high-symmetry principal facets, Fig.~\ref{fig:fig_1}~(a), of Pt, experiments give values $\phi$~(111)$= 5.6$-$6.4$~eV, $\phi$~$(110)= 5.4$-$5.7$~eV (reconstruction-dependent), and $\phi$~$(100)= 5.4$-$5.9$~eV, with measurement uncertainties $\sim0.1$~eV.~\cite{der15} These ranges of values are consistent with well-converged density functional theory (DFT) calculations.~\cite{sch21} Smoluchowski explained $\phi$ anisotropy to result from charge spreading and localization between the differing densities of surface atoms on crystal faces.~\cite{smo41} Recently, an extensive DFT study of this effect across many elements was published in ref.~[\onlinecite{tra19}]. Some other notable approaches to relate $\phi$ to varying crystal orientations include work by Brodie et al., that relates anisotropy to various other properties (effective mass, surface relaxation), and Fazylov, relating anisotropy to dielectric function and plasmon dispersions.~\cite{bro95,faz14} We find that Smoluchowski's ansatz is an intuitively appealing approach that reasonably estimates $\phi$ anisotropy.

To measure $\phi$ variation owing to crystalline anisotropy, Besocke et al. studied surfaces oriented between principal facets, using carefully machined macroscopic monocrystalline surfaces curving smoothly away from (111), Fig.~\ref{fig:fig_1}~(b). Besocke reported that $\phi$ decreased linearly by $\sim0.05$~eV/degree of surface slope for up to several degrees. Besocke explained the result using a Smoluchowski-type model in which surface atomic steps (inferred from electron diffraction) introduce electric dipoles ($\sim0.6$~D/edge atom) that counteract the overall surface dipole, thereby lowering $\phi$. Later, providing microscopic data substantiating Besocke’s model, Jia et al., and Park et al., used scanning probe methods to resolve individual step dipoles on Pt and other metals, but they did not provide a magnitude for the Pt(111) atomic step edge dipole moment.~\cite{Jia98,par05}

Here, we show how nanoscale surface roughness arising from polycrystalline textures impacts the local $\phi$ of sputter-deposited polycrystalline Pt thin films with (111) texture. Clean polycrystalline metal thin films with textured granular structure serve as a controlled and conceptually straightforward departure from Besocke's ideal oriented single crystals toward ensemble disorder by introducing roughness with varying small-angle surface slopes, $S$, and granular tilts, $T$ as depicted in Fig.~\ref{fig:fig_1}~(c). In effect, a sputter-deposited polycrystalline film serves as an ensemble of tiny monocrystalline samples, each with a few vicinal surfaces and varying atomic step densities that depend upon both local slope and granular tilts distributed around the mean surface normal, Fig.~\ref{fig:fig_1}~(c) and~(d).  Using spatially-resolved photoemission threshold mapping, we find a mean $\phi=5.70$~eV spread by $\pm0.03$~eV over a spatial distribution of many grain surfaces with random small-angle (few degree) (111)-vicinal orientations. Scanning tunneling microscopy (STM) indicates that the surfaces are predominately (111) step-terrace structure with widely-varying densities of atomic steps at numerous sites. Therefore, leveraging Besocke's result, we explain the spatially-varying work function using a Smoluchowski-smoothing model in which local $\phi$ decreases in direct proportion to the local density of atomic steps (electrical dipoles).~\cite{bes77} Since measuring atomic step densities over meaningfully large-area statistical samples of numerous micrometer-sized grains is not practical, we estimate atomic step distribution statistics from slopes, $S$, via atomic force microscopy (AFM) data, and tilts, $T$, via electron backscatter diffraction (EBSD) data. To estimate the atomic step dipole distributions resulting from convolved effects of $S$ and $T$, we find that a surprisingly straightforward estimate treating $S$ and $T$ as independent random variables leads to a reasonable description of the photoemission $\phi$ properties. Our model is a straightforward basis for understanding the measured spatially inhomogeneous $\phi$, while capturing the salient low $\phi$ values likely to be important for initial electron emission and simultaneously matching the narrow peak ($\pm0.03$~eV). The model estimates the step dipole strength to be $0.12$ D/atom, which is reasonable and similar to values from prior works.~\cite{bes77} We show that Smoluchowski's smoothing ansatz, substantiated by Besocke's mesocale studies of "ideal" specially-prepared surfaces and Jia and Park's microscopic atomic-step dipole measurements, create a basis to understand the impact of surface roughness on local $\phi$ for "realistic" surfaces prepared by commonplace industrial methods that yield ensemble disorder.

\section{\label{sec:level1} Experiment: materials synthesis and structure}
We sputter-deposited polycrystalline Pt (poly-Pt) thin film samples following the procedure in ref. [\onlinecite{she12}]. Samples are $\sim90$ nm-thick Pt on a $40$ nm-thick ZnO adhesion layer, on $400$ nm of thermal SiO$_{2}$ on Si~($100$). We annealed the samples for 1 hour in air at 900$^o$~C to create a coarse polycrystalline microstructure, with several-hundred-nanometer diameter grains as determined by AFM and transmission electron (TEM). 

\begin{figure*}
\includegraphics[width=0.9\textwidth]{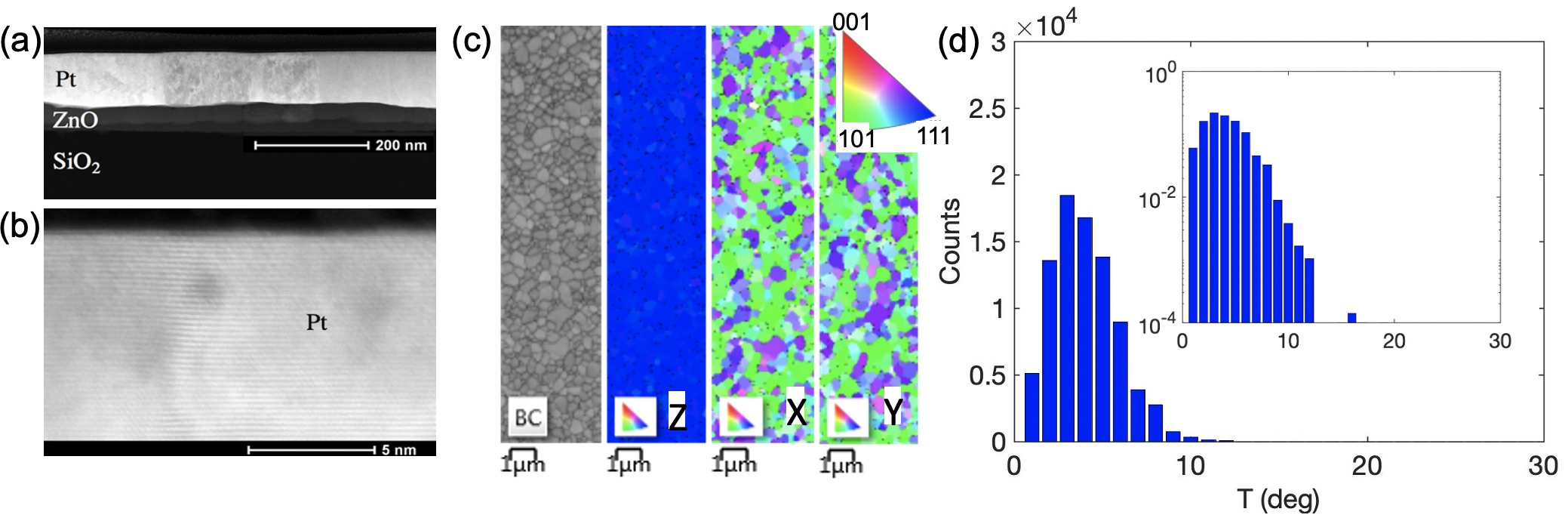}
\caption{\label{fig:fig_2} Grain structure and tilts of poly-Pt. (a) Cross-sectional TEM image of the thin film stack. The topmost layer in (a) is the Pt film. (b) High-resolution TEM image of the poly-Pt film. Periodic intensity in the image (along the horizontal direction of the image) indicates the crystallinity of Pt grains. (c) EBSD image based on Kikuchi band contrast displaying the boundaries between crystalline Pt grains and inverse pole figure maps (color) of orientations with respect to the sample substrate normal (Z) and two in-plane, orthogonal directions (X,Y). The color corresponds to the crystallographic orientations specified by the color key shown in the upper right. (d) Grain tilt histogram showing tilt with respect to the surface normal. The inset shows the histogram normalized to a probability density, $P_{T}(T)$, and plotted on a semilog scale to show the extent of the distribution. The units on the vertical axis are (degree)$^{-1}$.}
\end{figure*}

We begin by characterizing the structure, crystallographic orientation of poly-Pt grains, and topography of grain faces, since these properties will be used to explain $\phi$ spatial variation in a model presented in a subsequent section. Figure~\ref{fig:fig_2}~(a-b) shows cross-sectional TEM images of our sample. The top-most layer is the Pt film, with a ZnO adhesion layer and thermal SiO$_{2}$ layer beneath it. In Fig.~\ref{fig:fig_2}~(a), each Pt grain displays internal structures with slightly different intensities of grey. The high-resolution TEM image [Fig.~\ref{fig:fig_2}~(b)] shows periodic intensity variation due to electron beam diffraction from crystal planes, evincing the crystallinity of individual grains. Similar intensity variation observed across multiple grains suggests that they may be textured similarly. 

We measure ensemble distributions of grain crystal orientations (tilt) with respect to the substrate normal using EBSD. Figure~\ref{fig:fig_2}~(c) presents plan-view scanning electron microscopy (SEM) images (in gray scale) of poly-Pt films. Images based on band contrast show boundaries between crystalline Pt grains from which individual grains can be identified. Figure~\ref{fig:fig_2}~(c) (color panels Z,X,Y) are inverse pole figure maps of local crystallographic orientations with respect to the sample substrate normal (Z), and two in-plane, orthogonal directions (X,Y). The predominantly blue color with minimal variation in Z indicates that Pt grains are predominantly (111)-oriented. In Fig.~\ref{fig:fig_2}~(d), a histogram of surface-normal (Z) angles indicates a small median crystal tilt angle of $3.1^\circ$ with a standard deviation $1.5^\circ$. In-plane components of EBSD data (Fig.~\ref{fig:fig_2}~(c): X,Y), on the other hand, show the tendency for crystal domains to orient randomly in the azimuth. 

\begin{figure*}
\includegraphics[width=0.9\textwidth]{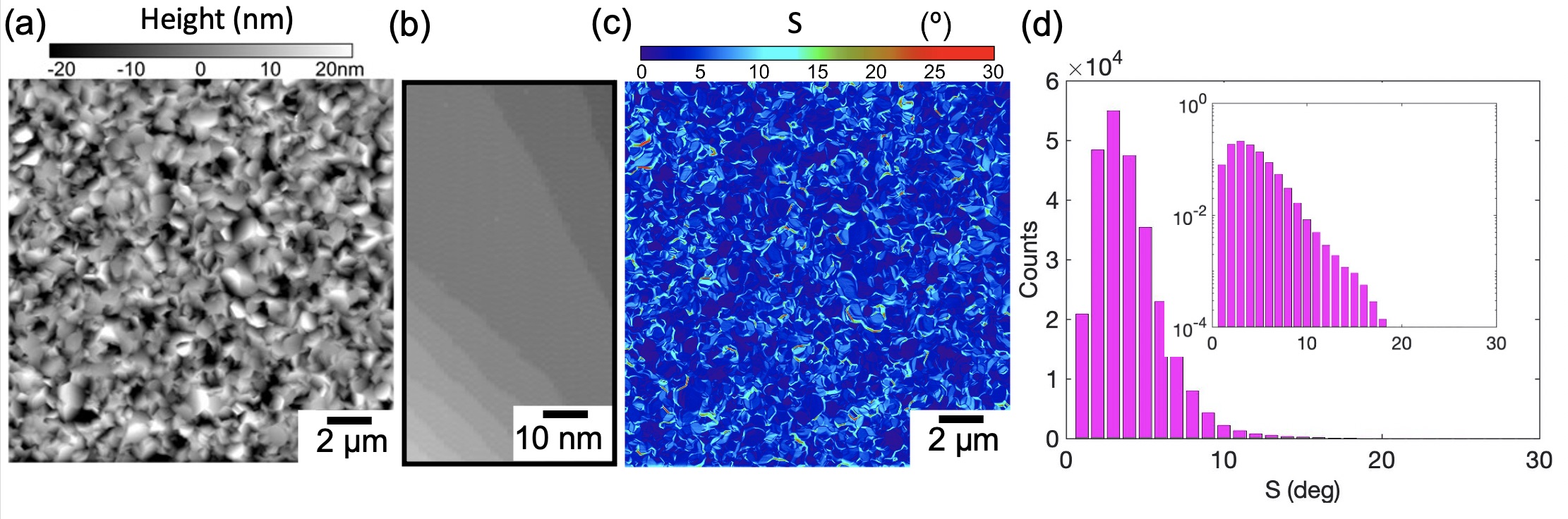}
\caption{\label{fig:fig_3} Poly-Pt surface topography via AFM and STM, along with processed images showing surface slope distribution. (a) Surface topography of poly-Pt, measured by AFM. (b) An example of the (111) atomic step structure over the 100-nm-scale resolved by STM. (c) A calculation of the local surface slope angle, $S$ (c) and a slope histogram (d). The inset shows the histogram normalized as a probability density, $P_{S}(S)$. The units on the vertical axis are (degree)$^{-1}$.}
\end{figure*}

To measure surface topography and slope, $S$, we performed AFM, as well as STM. Using STM, we resolved the surface step-terrace structure at up to the few-hundred-nanometer distances, while AFM is used to sample topography of grain ensembles over $\mu$m lengths. In Fig.~\ref{fig:fig_3}~(a), we show an AFM surface topography image containing many grains. At many locations on the surface, STM topography images, Fig.~\ref{fig:fig_3}~(b), reveal atomic step-terrace structure and step heights ($2.3$\AA) consistent with low-angle vicinal surfaces on Pt($111$) grains.~\cite{sti97} This observation is consistent with the AFM topography and the TEM data (Fig. 2). Prior experiments on Pt and numerous other fcc metals have shown that at small-angle ($\sim 1^{\circ}$) grain tilts and slopes near a high-symmetry direction, the surface is dominated by terraces of the nearby high-symmetry facet separated by single-height atomic steps.~\cite{bot92,jeo99} Consistent with the rolling topography in AFM, and grain tilts, the terrace size varies significantly with spatial location, as indicated even over the 10-nm-distances in Fig.~\ref{fig:fig_3}~(b). Figures~\ref{fig:fig_3}~(c) and (d) show a map and histogram of the slope angle, $S$, calculated from the AFM topography.  The distribution has a median of  $3.1^\circ$ and a standard deviation of $2.2^\circ$, underlining that a majority of surfaces are vicinal to (111), and unlikely to be other facets. The $S$ histogram is shown normalized to a probability density, $P_{S}(S)$, in the inset. The range of small slope angles is consistent with step-terrace structure in STM.

\section{\label{sec:level1} Experiment: photoemission work function maps}
\begin{figure}
\includegraphics[width=0.45\textwidth]{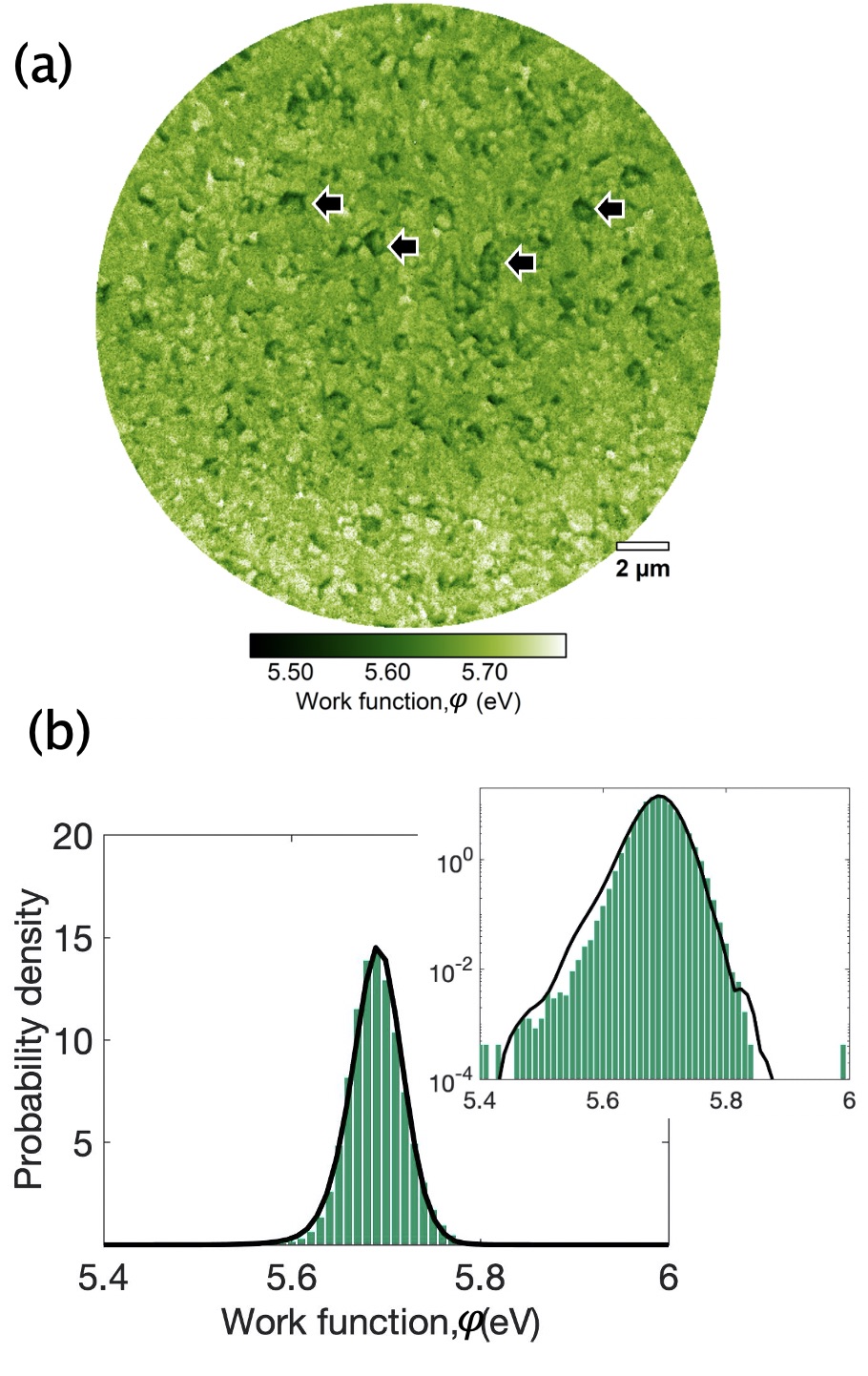}
\caption{\label{fig:fig_4} Photoemission threshold maps of poly-Pt. (a) Local $\phi$ map. The field of view is $24\mu$m. $\phi$ was obtained from local photoemission threshold, taking into account the Schottky effect originating from the instrument’s electric field. Black arrows denote pitted areas where Pt had been removed during annealing and that appear as low-$\phi$ areas.  (b) Histograms of work function values, normalized as a probability density, $P_{\phi}$, in which bin sizes correspond to uncertainties in measured values. The black solid line is a best-fit generated by our model described in the discussion. Inset: $P_{\phi}$ plotted on a semilog scale to emphasize the notable long-tail skew of the distribution. The units of the vertical axis are (eV)$^{-1}$.}
\end{figure}

The grain-to-grain $T$ variation (Fig.~\ref{fig:fig_2}~(c)), along with the variation of $S$ over granular vicinal surfaces (Fig.~\ref{fig:fig_3}~b)), is anticipated to produce spatially inhomogenous electronic property variations. To reveal the spatial $\phi$ distribution, we utilize photoemission electron microscopy (PEEM) equipped for spectroscopic imaging under ultrahigh vacuum ($\le10^{-9}$ Torr).~\cite{sha21}  A brief anneal ($900^{\circ}$, 3 minutes) in ultrahigh vacuum conditioned the sample surface for PEEM measurement. Figure~\ref{fig:fig_4} shows a PEEM $\phi$ map, as well as its distribution. The $\phi$ map, Fig.~\ref{fig:fig_4}~(a), varies spatially at sub-micron length scales, consistent with variation with grains and sub-grain vicinal faces.  Fig.~\ref{fig:fig_4}~(a) also shows some areas with relatively low $\phi$ (e.g. black arrows), which manifest after annealing in ultrahigh vacuum (UHV). AFM of the surface after PEEM measurement confirmed that these low $\phi$ regions corresponded to pits in the poly-Pt film, where Pt had been removed via dewetting effects.~\cite{jah14} Accordingly, we excluded pixels of dewetted areas from the PEEM $\phi$ distributions shown in Fig.~\ref{fig:fig_4}~(b).

The $\phi$ distribution has a mean of $5.70$~eV and a standard deviation of $\pm 0.03$~eV. To better estimate $\phi$, we accounted for the Schottky effect originating from the electric field $\sim10^7$~V/m, due to the measurement apparatus, at the sample surface, which lowers the surface barrier to emission by $\sim0.098$~eV.~\cite{sch14,ren06} For comparison, prior reports on precisely (111)-aligned single crystals give a range $\phi = 5.6-6.4$ eV using comparable photoemission and analysis techniques, with uncertainties of $\sim\pm 0.1$ eV. 

\section{\label{sec:level1} Discussion}
In relation to field emission, salient properties in Fig.~\ref{fig:fig_4}, are:  (1) that the submicron-scale textures evident in the $\phi$ spatial map, Fig.~\ref{fig:fig_4} ~(a), indicate that $\phi$ varies spatially in connection with varying local surface properties, and (2) that the $\phi$ distribution is skewed with a tail toward low $\phi$ far below the mean value. In this section, we develop a model that simultaneously approximates both the $\phi$-spatial variation in connection with surface structure, and the long-tail skew distribution to low $\phi$-values. 

Motivated by Besocke et al.'s work connecting mesoscale local $\phi$ with local surface atomic step density, substantiated by microscopic observations of individual atomic-step dipoles on Pt(111) and several other fcc metals, we utilize a Smoluchowski-type model of local $\phi$, relative $\phi$~(hkl) (here for flat Pt(111)) to approximate $\phi$ spatial variation due to surface morphology.~\cite{bes77,par05,ren06} Our model expresses the microscopic viewpoint that work function variation $\Delta\phi =\phi-\phi$~(hkl) is proportional to the local atomic step density, $N$, and the strength of the atomic step dipole, $\mu$. Besocke’s expression relating local $\phi$ of a stepped vicinal surface to $\phi$(hkl) of the nearby high-symmetry plane is  $\Delta\phi = -300\times10^{-18}$~$4$~ $\pi$~$\mu$~$N$, in units of eV, where $N$ and $\mu$ are expressed in units cm$^{-1}$ and Debye (D)/cm, respectively.~\cite{bes77} The local atomic-step density, $N(S,T)=\sin(T-S)/a_{111}$, on a surface region with local tilt $T$ and slope $S$. Here, $a_{111}=a_{\circ}/\sqrt{3}=0.226$~nm is the distance between (111) planes, with Pt's fcc lattice constant $a_{\circ}=0.392$~nm. At small $S$ and $T$ angles as in our measurements, it is very accurate ($\lesssim$ 10$\%$ errors) to use the small-angle approximation, i.e. that N$\simeq(T - S)/a_{111}$.

A key motivation for our overall approach is that STM images, Fig.~\ref{fig:fig_3}~(b), indicate that many sites on our surfaces have distinctly resolved, smoothly varying, (111) step-terrace structure. Practical experimental speed limits restrict STM to mapping several sites with $\sim$100-nm-scale images, i.e. only providing exact descriptions of faces on several individual grains. 

To approximate mesoscale ensemble properties across many grains, we estimate the atomic-step density {\it distribution}, $N$, by computing it from from the slope, $P_{S}$, and tilt, $P_{T}$, distributions (panel (d) of Figs.~\ref{fig:fig_2} and \ref{fig:fig_3}). To do so, we must make a choice of how to estimate the random variable $T-S$. Although $S$ and $T$ have known underlying driving forces (e.g. surface energy anisotropy) that favor correlation (facets), we make the approximation that $S$ and $T$ are random, independent variables. Effectively, we are assuming that surface energy anisotropy is not so large (relative to $k_{B} T$) as to drive well-developed step-free facets, which is consistent with the observation of densely-stepped low-angle (vs. zero-angle) vicinal surfaces in many STM images (see Fig.~\ref{fig:fig_3}~(b)). In the approximation that $S$ and $T$ are uncorrelated random variables, the difference $d =T-S$ is calculated by discrete convolution, and the atomic step density distribution is

 \[N(d)=d/a_{111} = a_{111}^{-1}\sum_{k=-\infty}^{\infty} P_{S}(k)P_{T}(d+k)\],

\noindent such that,

\[\Delta\phi = -300\times10^{-18} 4 \pi \mu N \]

\[=-300\times10^{-18} 4 \pi \mu a_{111}^{-1} \sum_{k=-\infty}^{\infty} P_{S}(k)P_{T}(d+k),\]

\noindent where $P_{S}(S)$ and $P_{T}(T)$ are the measured probability densities for $S$ and $T$, and $k$ ranges over all values of $T$ and $S$. 

We fit our measured $\phi$ distribution using the $\Delta\phi$ expression above using $\mu$ as the fit parameter to match the distribution width and $\phi(111)$ to set the center position. The fit is shown in Fig.~\ref{fig:fig_4} by a solid black line. Both qualitative and quantitative features of the $\phi$ distribution are reasonably captured by the fit. First, and most significant for field emission, both the spatially patchy grain-to-grain $\phi$-variation evident in Fig.~\ref{fig:fig_4}~(a), and the peculiar $\phi$ distribution, Fig. 4 (b), with the relatively narrow $\sim0.03$~eV "head" and skewed long-tail to much lower $\phi\sim5.4$ eV can be understood to directly correspond to the spread of atomic step densities anticipated from the skewed $P_{S}$ and $P_{T}$ distributions. The best-fit values for the distributions are $\mu=0.12$ D/atom ($\mu\simeq 4\times 10^{6}$~D/cm$^{-1}$, weakly dependent upon step orientation/atomic density), and $\phi(111) = 5.69$ eV. 

Our atomic step dipole is lower than the value ($0.6\pm 0.1$~D/edge atom) determined by Besocke et al., using very different material preparations and macroscopic area-averaging measurement methods.~\cite{bes77} It is worth pointing out that Besocke's $\mu$ value is too large to reconcile with other data, e.g. for just $S\sim7^{\circ}$ slope it gives $\Delta\phi\sim0.3$~eV reduction in work function, which is comparable to reported variations between principal high-symmetry facets with a range $\sim 0.5$ eV between (110) and (111) facets. Our result ($\sim 0.06$ eV for $7^{\circ}$ from (111)) is more straightforward to reconcile within the $0.5$~eV full range of facet-to-facet work function anisotropy. Our value also compares reasonably in a broader context of various results obtained for other materials, which have yielded fairly wide-ranging values for step-edge dipoles for most fcc metals.~\cite{Jia98,bes77,par05} For example on Au(111), Jia et al. found a step dipole of 0.16 D/edge atom, while Besocke reports a value of $0.2$-$0.27$ D/edge atom, and Park measure 0.45 D/edge atom.~\cite{Jia98,bes77,par05} The step dipole strength produced by our model fit is reasonable in comparison to those reported values.

\section{\label{sec:level1} Summary}
We report and analyze $\phi$ for (111)-textured polycrystalline Pt thin films with grain-to-grain ensemble structural disorder. We measure a mean $\phi = 5.70\pm 0.03$ eV, which is consistent with reported $\phi = 5.6$-$6.4$~eV for intentionally aligned (111) facets. To explain the spatially varying, skewed $\phi$, we develop a Smoluchowski model that accounts for the influence of atomic step edge dipoles $\mu$. Connecting work function with step dipoles is motivated by prior studies indicating that Pt, like many other fcc metals, has step dipoles, and our observation of spatially varying vicinal (111) atomic step densities in STM images of our surfaces. Our analysis yields a Pt(111) atomic step dipole $\mu=0.12$~D/edge atom. Toward models of field emission and related phenomena, e.g. vacuum arc initiation, that employ realistic, rough, polycrystalline electrodes, two salient traits of our $\phi$ map and model are: (1) $\phi$ spatial variations in connection with surface grain/facet structure, and (2) skew to lower $\phi$ values down to 5.4 eV. Our analysis provides a good explanation of the granular ensemble $\phi$ properties, including the spatial variation and long-tail skew to lower $\phi$. Future work should include coincident-site measurements to directly correlate between local grain tilts (T), film topographies (S), and local $\phi$.


\begin{acknowledgments}
This work is funded by Sandia Laboratory Directed R\&D (LDRD) program and also by an Air Force Office of Scientific Research (AFOSR) program. Part of the work is performed at the Center for Integrated Nanotechnologies, a U.S. Department of Energy (DOE) Office of Science User Facility. Sandia National Laboratories is a multi-mission laboratory managed and operated by National Technology and Engineering Solutions of Sandia, LLC., a wholly owned subsidiary of Honeywell International, Inc., for the U.S. Department of Energy’s National Nuclear Security Administration under contract DE-NA0003525. This paper describes objective technical results and analysis. Any subjective views or opinions that might be expressed in the paper do not necessarily represent the views of the U.S. Department of Energy or the United States Government.

\end{acknowledgments}

\section*{Data Availability Statement}

The data that support the findings of this study are available within the article.

\section*{Conflict of Interest Statement}

The authors have no conflicts to disclose.

\bibliography{PTFE_v6.bib}

\end{document}